\documentstyle[a4wide,epsf]{article}
\begin{document}

\newcommand{\lsim}{\mbox{\raisebox{-.9ex}{~$\stackrel{\mbox{$<$}}{\sim}$~}}}
\newcommand{\gsim}{\mbox{\raisebox{-.9ex}{~$\stackrel{\mbox{$>$}}{\sim}$~}}}

\begin{center}
{\large\bf Hybrid Dark Sector: Locked Quintessence and Dark Matter}

\bigskip

{\large  Minos Axenides$^*$ and Konstantinos Dimopoulos$^*$}

\bigskip

$^*${\it 
Institute of Nuclear Physics, National Center for Scientific Research
`Demokritos',\\ 
Agia Paraskevi Attikis, Athens 153 10, Greece}

\begin{abstract}
We present a unified model of dark matter and dark energy. The dark matter 
field is a modulus corresponding to a flat direction of 
supersymmetry, which couples, in a hybrid type potential, with the dark 
energy field. The latter is a light scalar, whose direction is stabilized by 
non-renormalizable terms. This quintessence field is kept `locked' on top of 
a false vacuum due to the coupling with the oscillating dark matter field.
It is shown that the model can satisfy the observations when we consider
low-scale gauge-mediated supersymmetry breaking. The necessary initial 
conditions are naturally attained by the action of supergravity corrections 
on the potential, in the period following the end of primordial inflation.
\end{abstract}

\end{center}

\section{Introduction}

In the last few years cosmology experienced a dramatic influx of observational data \cite{wmap, sdss, 2dF, SN}. 
Analysis of these observations have resulted in the emergence of the
so-called `concordance' model of cosmology. According to the data
we live in a spatially flat Universe, whose content is comprised of
predominantly unknown substances, not accounted for by the standard model
of particle physics. In particular, roughly 1/3 of the energy density of our 
Universe behaves as pressureless matter (dust), with little or no interactions 
with the usual baryonic matter. These weakly interacting massive particles
(WIMPs) do not correspond to the luminous matter of galaxies and, hence, have
been named dark matter. The remaining 2/3 of the Universe content at present, 
is attributed to an even more exotic substance, whose properties are such that 
affect the global geometry of the Universe and cause the observed current 
accelerated expansion of spacetime. In analogy with dark matter, this 
substance has been named dark energy. Therefore, despite the substantial 
progress of particle cosmology in the last decade, cosmologists are forced to 
admit that the bulk of the content of the Universe corresponds to an unknown 
dark sector, on whose origin and theoretical justification we can only 
speculate.

Modern particle physics indeed offers a number of possibilities for
the explanation of this dark sector. In particular, supersymmetric theories 
include a zoo of unobserved particles corresponding to the superpartners of
standard model particles. This so-called hidden sector may soon be accessible 
to collider experiments. Other candidates are offered by modified gravity 
theories (e.g. the Brans-Dicke scalar of scalar tensor gravity), string 
theories (e.g. string axions/moduli, Kaluza-Klein particles) and theories of 
large extra dimensions (e.g. the radion). Therefore, it seems that particle 
physics has available options capable of addressing the problem of the dark 
sector.

Since the dark matter issue has been present for some time, a number of 
successful candidate WIMPs exist to explain it, the most prominent of which 
are the lightest supersymmetric particle (e.g. the neutralino) and the 
axion (i.e. the phase--field of the Peccei-Quinn field, used to solve the 
strong CP--problem). On the other hand, the problem of dark energy is quite 
recent and more difficult to address, because the properties of dark energy 
are quite bizarre and may even threaten some of our `fundamental prejudices' 
such as the vanishing of the cosmological constant or even the dominant energy 
condition (for a review see \cite{review}). 

The simplest form of dark energy is a non-zero cosmological constant $\Lambda$.
Phenomenologically, this is the most appealing choice, since it provides a very
nice fit to the data (the so-called $\Lambda$CDM model, whose minimal form is
the famous `vanilla model'). However, the value of the cosmological constant 
has to be fine tuned to the incredible level of 
\mbox{$\Lambda\sim 10^{-123}M_P^2$}, compared to its natural value, given by 
the Planck mass $M_P$. Moreover, a constant non-zero vacuum density inevitably 
leads to eternal accelerated expansion. This results in the presence of future 
causal horizons, which inhibit the construction of the S-matrix in string 
theory and are, therefore, most undesirable \cite{horizons}. 

For these reasons theorists have attempted 
to formulate alternative solutions to the dark energy problem, while keeping 
\mbox{$\Lambda=0$} as originally conjectured. The most celebrated such idea
is the introduction of the so-called quintessence field \cite{quint}; the fifth element 
after cold dark matter (WIMPs), hot dark matter (neutrinos), baryons and 
photons. Quintessence is a light scalar field $Q$, which has not yet reached  
the minimum of its potential and, therefore, is responsible for the presence
of a non-vanishing potential density $V_0$ today. This density currently
dominates the Universe, giving rise to an effective cosmological constant 
\mbox{$\Lambda_{\rm eff}=8\pi GV_0$}, which causes the observed
accelerated expansion. Eventually, the quintessence field will reach the 
minimum of its potential (corresponding to the true vacuum) ending, thereby, 
the accelerated expansion. Hence, quintessence dispenses with the future 
horizon problem of $\Lambda$CDM.

However, despite its advantages, the quintessence idea suffers from certain 
generic problems \cite{KL}. For example, in order to achieve the correct value 
of $V_0$, one usually needs to fine-tune accordingly the 
quintessence potential. Also, in fairly general grounds it can be shown that 
the value of quintessence at present is \mbox{$Q\sim M_P$} (if originally at 
zero) with a tiny effective mass \mbox{$m_Q\sim 10^{-33}$eV}. In the context 
of supergravity 
theories such a light field is difficult to understand because the flatness of 
its potential is lifted by excessive supergravity corrections or due to the 
action of non-renormalizable terms, which become important at displacements 
of order $M_P$. Finally, quintessence introduces a second tuning problem, 
that of its initial conditions.

In this paper we attempt to address the dark sector problem in a single
theoretical framework. Other such attempts can be found in Ref.~\cite{other}.
We assume that the dark matter particle is a modulus
$\Phi$, corresponding to a flat direction of supersymmetry. The modulus field
is undergoing coherent oscillations, which are equivalent to a collection of
massive $\Phi$--particles, that are the required WIMPs. Coupled to the dark
matter is another scalar field $\phi$. This can be thought of as our 
quintessence
field and it corresponds to a flat direction lifted by non-renormalizable 
terms. Even though the $\phi$--field is a light scalar, it is much more 
massive than the $m_Q$ mentioned above, so as not to be in danger from 
supergravity corrections to its potential. Our quintessence field is coupled
to our dark matter in a hybrid manner, which is quite natural in the context
of a supersymmetric theory. Due to this coupling, the oscillating $\Phi$,
keeps $\phi$ `locked' on top of a potential hill, giving rise to the desired
dark energy. When the amplitude of the $\Phi$--oscillations decreases enough,
the dark energy dominates the Universe, causing the observed accelerated 
expansion. Much later, when the oscillation amplitude is reduced even further,
the `locked' quintessence field is released and rolls down to its minimum.
Then, the system reaches the true vacuum and accelerated expansion ceases.
Our model accounts successfully for the observations using natural mass-scales
(corresponding to low-scale gauge-mediated supersymmetry 
breaking). In order to explain the required initial conditions we explore in
detail the history of our system during the early Universe, when the 
supergravity corrections to the scalar potential are essential.

Our paper is organized as follows. In Sec.~2 we present and analyze the 
dynamics of our model, while we also determine the value of the model 
parameters. In Sec.~3 we demonstrate that the required initial conditions for
our dark matter field $\Phi$ may be naturally attained due to the action of 
supergravity corrections on the scalar potential. 
We also investigate the 
disastrous possibility of the decay of the oscillating dark matter condensate 
into quintessence quanta. 
In Sec.~4 we show that 
the supergravity corrections may also ensure the locking of the quintessence 
field $\phi$. Additionally, we elaborate more on the value of the tachyonic 
mass of $\phi$ and its vacuum expectation value. Finally, in Sec.~5 we discuss 
our results and present our conclusions.

We assume a spatially flat Universe, according to the WMAP observations 
\cite{wmap}.
Throughout our paper we use natural units such that \mbox{$\hbar=c=1$} and
Newton's gravitational constant is \mbox{$8\pi G=m_P^{-2}$}, where 
\mbox{$m_P=2.4\times 10^{18}$GeV} is the reduced Planck mass.

\section{The model}\label{model}

Consider two real scalar fields $\Phi$ and $\phi$ with a
hybrid type of potential of the form
\begin{equation}
V(\Phi,\phi)=\frac{1}{2}m_\Phi^2\Phi^2+
\frac{1}{2}\lambda\Phi^2\phi^2+\frac{1}{4}\alpha(\phi^2-M^2)^2,
\label{V}
\end{equation}
where \mbox{$\lambda\lsim 1$}.
From the above we see that the tachyonic mass of $\phi$ is given by
\begin{equation}
m_\phi=\sqrt{\alpha}\,M.
\label{mphi0}
\end{equation}
The above potential has global minima at \mbox{$(\Phi,\phi)=(0,\pm M)$} and
an unstable saddle point at \mbox{$(\Phi,\phi)=(0,0)$}.
Now, since the effective mass--squared of $\phi$ is
\begin{equation}
(m_\phi^{\rm eff})^2=\lambda\Phi^2-\alpha M^2,
\label{mphi}
\end{equation}
if \mbox{$\Phi>\Phi_c$} then $\phi$ is driven to zero,
where 
\begin{equation}
\Phi_c\equiv\sqrt{\frac{\alpha}{\lambda}}\;M.
\label{Phic}
\end{equation}

Suppose that originally the system lies in the regime, where,
\mbox{$\Phi\gg\Phi_c$} and \mbox{$\phi\simeq 0$}. With such initial 
conditions the effective potential for $\Phi$ becomes quadratic:
\begin{equation}
V(\Phi,\phi=0)=\frac{1}{2}m_\Phi^2\Phi^2+V_0.
\label{over}
\end{equation}
Hence, when $\phi$ remains at the origin, $\Phi$ oscillates on top of a 
false vacuum with density
\begin{equation}
V_0=\frac{1}{4}\,\alpha M^4.
\label{V0}
\end{equation}
The oscillation frequency is \mbox{$\omega_\Phi\sim m_\Phi$} and the time 
interval $(\Delta t)_s$ that the field spends on top of the saddle point 
(\mbox{$\Delta\Phi\leq\Phi_c$}) is
\begin{eqnarray}
\omega_\Phi\Delta t\sim\frac{\Delta\Phi}{\bar{\Phi}} & \Rightarrow & 
(\Delta t)_s\sim\frac{\Phi_c}{m_\Phi\bar{\Phi}}
\label{Dt}
\end{eqnarray}
where $\bar{\Phi}$ is the amplitude of the oscillations. Originally this 
amplitude may be quite large but the expansion of 
the Universe dilutes the energy of the oscillations and, therefore, 
$\bar{\Phi}$ decreases, which means that $(\Delta t)_s$ grows. 

However, as long as the system spends most of this time away from the saddle
and until $(\Delta t)_s$ becomes large enough to be comparable to the inverse 
of the tachyonic mass of $\phi$, the latter has no time to roll away from the 
saddle \cite{dvali,ours}. Hence, the oscillations of $\Phi$ on top of the 
saddle can, in principle, continue until the amplitude decreases down to
\begin{equation}
\Phi_s\sim\frac{\Phi_c\,m_\phi}{m_\Phi}
\sim\frac{\alpha M^2}{\sqrt{\lambda}\,m_\Phi}
\label{Phis}
\end{equation}
at which point $\phi$ has to depart from the origin and roll down toward its 
vacuum expectation value (VEV) $M$. However, the roll down of $\phi$ can occur
earlier if \mbox{$\Phi_c>\Phi_s$}, even though the period of oscillation is 
smaller than $m_\phi^{-1}$. Indeed, when \mbox{$\Phi_s<\bar{\Phi}<\Phi_c$}, 
\mbox{$\phi\simeq 0$} is not possible because, were it otherwise, it would 
mean that the field would have had to remain on top of the saddle for the 
entire period of oscillation. Hence, $\phi$ departs from the origin at 
$\bar{\Phi}_{\rm end}$, where
\begin{equation}
\bar{\Phi}_{\rm end}\equiv{\rm max}\{\Phi_c, \Phi_s\}\,.
\label{Phiend}
\end{equation}
From Eqs.~(\ref{Phic}) and (\ref{Phis}) we find that $\bar{\Phi}_{\rm end}$ 
is decided by the relative magnitude of the masses of the scalar fields because
\begin{equation}
\frac{\Phi_c}{\Phi_s}\sim\frac{m_\Phi}{m_\phi}\,.
\label{Phiratio}
\end{equation}

During the oscillations the density of the oscillating $\Phi$ is 
\begin{equation}
\rho_\Phi=\frac{1}{2}\dot{\Phi}^2+\frac{1}{2}m_\Phi^2\Phi^2\simeq
\frac{1}{2}m_\Phi^2\bar{\Phi}^2,
\label{rPhi}
\end{equation}
where the dot denotes derivative with respect to the cosmic time $t$.
Comparing this with the overall potential density given in Eq.~(\ref{over}) we 
see that, the overall density is dominated by the false vacuum density given 
in Eq.~(\ref{V0}), when the oscillation amplitude is smaller than, 
\begin{equation}
\Phi_\Lambda
\sim\frac{\sqrt{\alpha}\,M^2}{m_\Phi}
\sim\left(\frac{m_\phi}{m_\Phi}\right)M\,.
\label{PhiL}
\end{equation}

The above model can be used to account for both the dark matter and the dark
energy in the Universe. Provided the initial conditions for the two fields
are appropriate, it is possible that the oscillating field $\Phi$ constitutes
the dark matter, whereas the field $\phi$ is responsible for eliminating
the dark energy in the future, so as to avoid eternal acceleration and future 
horizons. The dark matter field $\Phi$ oscillates on top of the false vacuum
$V_0$ in the same manner as in `locked' inflation \cite{dvali,ours}.
The false vacuum is not felt until today, when the accelerated expansion 
begins. Eventually, at some moment in the future, the amplitude of the 
$\Phi$--oscillation reaches $\bar{\Phi}_{\rm end}$ and $\phi$ rolls away from 
the origin terminating the accelerated expansion.

We want our model to explain the dark energy responsible for the currently 
observed accelerated expansion of the Universe. Hence, the false vacuum 
density $V_0$ of our model should be comparable to the density $\rho_0$ of 
the Universe at present
\begin{equation}
V_0\sim\rho_0\sim 10^{-120}m_P^4\sim (10^{-3}{\rm eV})^4.
\label{r0}
\end{equation}
In view of Eq.~(\ref{V0}), this implies the condition
\begin{equation}
M\sim\alpha^{-1/4}10^{-30}m_P\;.
\label{Mcond}
\end{equation}

We also want our model to explain the dark matter, by means of the oscillating
scalar field $\Phi$. Indeed, it is well known that a scalar field oscillating
in a quadratic potential has the equation of state of pressureless matter
\cite{turner} (it corresponds to a collection of massive $\Phi$--particles)
and, therefore, $\Phi$ can account for the dark matter necessary
to explain the observations. For this, the oscillating $\Phi$ has to satisfy 
certain requirements. One of these is the obvious requirement that $\Phi$ 
should not have decayed until today. This means that the decay rate of $\Phi$
should satisfy the condition
\begin{equation}
\Gamma_\Phi<H_0\;,
\label{GPhi}
\end{equation}
where \mbox{$H_0\sim\sqrt{\rho_0}/m_P$} is the Hubble parameter at present.
Using that \mbox{$\Gamma_\Phi\sim g_\Phi^2m_\Phi$} we find the bound
\begin{equation}
m_\Phi\leq 10^{-20}m_P\;,
\label{mPhibound}
\end{equation}
where we used that the coupling $g_\Phi$ of $\Phi$ with its decay products 
lies in the range \mbox{$\frac{m_\Phi}{m_P}\leq g_\Phi\leq 1$}, with the lower 
bound corresponding to the gravitational decay of $\Phi$, for which 
\mbox{$\Gamma_\Phi\sim m_\Phi^3/m_P^2$}. From the above bound we see that we
require $\Phi$ to be a rather light field with mass \mbox{$\lsim$ 10 MeV}. 

We choose, therefore, to use a modulus field, corresponding to a flat direction
of supersymmetry, whose mass is estimated as
\begin{equation}
m_\Phi\sim\frac{M_S^2}{m_P}\,,
\label{mPhi}
\end{equation}
where $M_S$ is the supersymmetry breaking scale, ranging between
\mbox{$m_{3/2}\leq M_S\leq\sqrt{m_Pm_{3/2}}$}, where \mbox{$m_{3/2}\sim$ 1 TeV}
is the electroweak scale (gravitino mass) and the upper bound corresponds to
gravity mediated supersymmetry breaking while the lower bound corresponds to 
gauge mediated supersymmetry breaking, which can give $M_S$ as low as
\mbox{(few) $\times$ TeV}. Eqs.~(\ref{mPhibound}) and 
(\ref{mPhi}) suggest
\begin{equation}
1\;{\rm TeV}\leq M_S\leq 10^{-10}m_P\;.
\label{MSrange}
\end{equation}

\begin{center}
\begin{figure}
\label{fig}

\begin{center}
\leavevmode
\hbox{%
\epsfxsize=3in
\epsffile{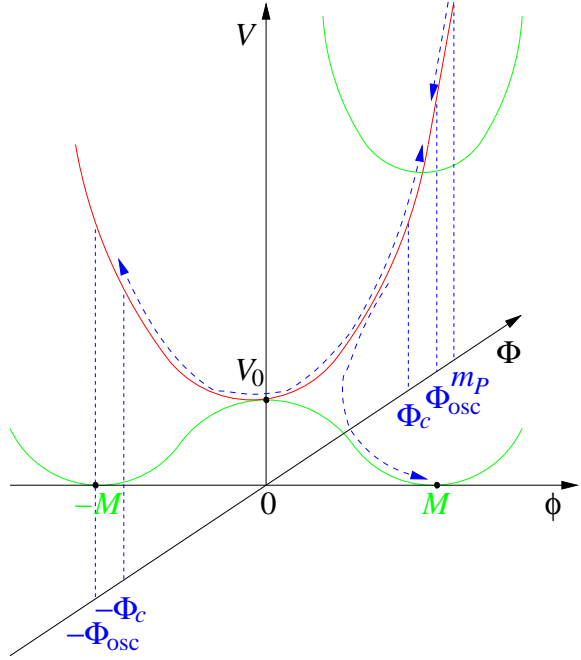}}
\end{center}

\caption{\footnotesize 
Illustration of the scalar potential $V(\Phi,\phi)$. Originally, 
\mbox{$\Phi\sim m_P$} and \mbox{$\phi\simeq 0$}. The field $\Phi$ begins
oscillating with amplitude $\Phi_{\rm osc}$. The Universe dilutes the energy
of the oscillations until the amplitude decreases to 
\mbox{$\bar{\Phi}_{\rm end}\sim\Phi_c$}, when the system departs from the 
saddle and rolls toward the minimum at \mbox{$(\Phi=0, \phi=\pm M)$}.
}
\end{figure}
\end{center}

If $\phi$ were a modulus too then the natural value of $\alpha$ would have been
\begin{equation}
\alpha\sim\left(\frac{M_S}{m_P}\right)^4,
\label{alpha}
\end{equation}
with \mbox{$M\sim m_P$}. The above, however, in view of Eq.~(\ref{Mcond}), 
results in the condition
\begin{equation}
M\sim 10^{-30}\frac{m_P^2}{M_S}\,,
\label{MMScond}
\end{equation}
which, combined with the range in Eq.~(\ref{MSrange}), results in the
following range for the vacuum expectation value (VEV) of $\phi$
\begin{equation}
10\;{\rm MeV}\leq M\leq 1\;{\rm TeV}\,.
\label{Mrange}
\end{equation}
Thus, we see that \mbox{$M\ll m_P$} in contrast to expectations.
However, there are ways to reduce the VEV of $\phi$, provided the
tachyonic mass $m_\phi$ remains roughly unmodified. Hence, in the following we
retain the value of $\alpha$ shown in Eq.~(\ref{alpha}).
We discuss the small VEV of $\phi$ in Sec.~\ref{VEV}.
In view of Eqs.~(\ref{MSrange}) and (\ref{Mrange}) we make the following choice
\begin{equation}
M\sim M_S\sim 1\;{\rm TeV}.
\label{choice}
\end{equation}
With this choice, the number of parameters of the model in Eq.~(\ref{V}) is 
minimized to two natural mass scales: $m_P$ and \mbox{$M_S\sim m_{3/2}$} and a 
coupling \mbox{$\lambda\leq 1$}. 

Before concluding this section we notice that, from Eqs.~(\ref{mphi}),
(\ref{alpha}) and (\ref{choice}) we find
\begin{equation}
m_\phi\sim\frac{M_S^3}{m_P^2}\,,
\label{mphiMS}
\end{equation}
which, in view of Eqs.~(\ref{Phiend}) and (\ref{Phiratio}), gives
\begin{equation}
\frac{\Phi_c}{\Phi_s}\sim\frac{m_P}{M_S}\gg 1 \quad\Rightarrow\quad
\bar{\Phi}_{\rm end}\sim\Phi_c\;.
\label{PhiratioPhiend}
\end{equation}
Now, since we need the oscillation of $\Phi$ on top of the false 
vacuum to continue until today, when \mbox{$\bar{\Phi}\sim\Phi_\Lambda$}, we
require
\begin{equation}
\frac{\Phi_\Lambda}{\bar{\Phi}_{\rm end}}\sim\sqrt{\lambda}
\left(\frac{m_P}{M_S}\right)>1 \quad\Rightarrow\quad
\lambda>10^{-30}
\label{lrange}
\end{equation}
where we also used Eqs.~(\ref{Phic}) and (\ref{PhiL}).

\section{Dark matter requirements}

\subsection{\boldmath The value of $\Phi$ at the onset of the oscillations}

Another important requirement for $\Phi$, if the latter is to account for the
dark matter in the Universe, is that it has the correct energy density. This
requirement is determined by the initial amplitude $\Phi_{\rm osc}$ of the 
oscillations of the field. When the oscillations begin we have 
[cf. Eq.~(\ref{rPhi})]
\begin{equation}
\rho_\Phi^{\rm osc}=\frac{1}{2}m_\Phi\Phi_{\rm osc}^2\;,
\label{rPhiosc}
\end{equation}
where the subscript `osc' denotes the onset of the $\Phi$--oscillations. 
According to the Friedman equation we have \mbox{$\rho=3H^2m_P^2$}, where 
\mbox{$H(t)\equiv\dot{a}/a$} is the Hubble parameter and $a(t)$ is the scale 
factor, parameterizing  the Universe expansion. Hence, since the 
oscillations begin when \mbox{$H_{\rm osc}\sim m_\Phi$}, we find
\begin{equation}
\left.\frac{\rho_\Phi}{\rho}\right|_{\rm osc}\sim
\left(\frac{\Phi_{\rm osc}}{m_P}\right)^2.
\label{ratioosc}
\end{equation}
Now, using Eq.~(\ref{mPhi}) and also that, during the radiation era, 
\mbox{$\rho\sim T^4$} (with $T$ being the temperature) we obtain
\begin{equation}
\frac{m_\Phi}{H_{\rm eq}}\sim\left(\frac{M_S}{T_{\rm eq}}\right)^2\gg 1
\end{equation}
where `eq' denotes the time $t_{\rm eq}$ of equal matter and radiation 
densities, at which \mbox{$T_{\rm eq}\sim 1$ eV}. Hence, we see that 
\mbox{$H_{\rm osc}\gg H_{\rm eq}$}, which means that the oscillations begin 
during the radiation dominated period. During this period the density of the
Universe scales as \mbox{$\rho\propto a^{-4}$}, while the density of the 
oscillating scalar field scales as \mbox{$\rho_\Phi\propto a^{-3}$} 
\cite{turner}. Hence we have \mbox{$\rho_\Phi/\rho\propto a\propto H^{-1/2}$}.
Therefore, the density of the oscillating scalar field eventually dominates
the Universe. Since, we want $\Phi$ to be the dark matter, we require that
its density dominates at $t_{\rm eq}$. Consequently, Eq.~(\ref{ratioosc}) 
suggests
\begin{equation}
\left.\frac{\rho_\Phi}{\rho}\right|_{\rm osc}\sim
\sqrt{\frac{H_{\rm eq}}{m_\Phi}}
\quad\Rightarrow\quad\Phi_{\rm osc}\sim
\left(\frac{T_{\rm eq}^2}{m_\Phi m_P}\right)^{1/4}m_P
\sim\sqrt{\frac{T_{\rm eq}}{M_S}}\;m_P\;,
\label{Phiosc}
\end{equation}
where we also used Eq.~(\ref{mPhi}). Putting the numbers in the above we find
\mbox{$\Phi_{\rm osc}\sim 10^{-6}m_P$}. This is substantially smaller than the
natural expectation for a modulus, which corresponds to an original 
misalignment (i.e. displacement from its VEV) of order $m_P$. However, below 
we attempt to explain this reduced misalignment by means of supergravity 
corrections. These corrections are expected to lift the flatness of the 
$\Phi$-direction and enable $\Phi$ to begin rolling down long before
\mbox{$H\sim m_\Phi$}.

\subsection{The effect of supergravity corrections}

Supergravity corrections to the potential generate an effective mass term
proportional to the Hubble parameter \cite{sugra}. Thus, the effective 
potential in Eq.~(\ref{over}) becomes
\begin{equation}
V(\Phi)=\frac{1}{2}[\pm cH^2(t)+m_\Phi^2]\Phi^2,
\label{Vsugra}
\end{equation}
where $c$ is a positive constant and we ignored the false vacuum contribution 
$V_0$, which is negligible at times much earlier than the present time. 

We assume that, in the early stages of its evolution, the Universe underwent
a period of cosmic inflation. 
%
During and after inflation, until reheating, the Universe is dominated by the
density of the inflaton field. The minimum of $V(\Phi)$, in general, is 
expected to be shifted by \mbox{$\Delta\Phi\sim m_P$} at the end of 
inflation.
Hence, at the end of inflation we expect \mbox{$\Phi_{\rm inf}\sim m_P$}.
After the end of inflation and until reheating $V(\Phi)$ is given by
Eq.~(\ref{Vsugra}) with \mbox{$c\sim{\cal O}(1)$} \cite{sugra}. 
However, after reheating, when the Universe becomes radiation dominated,
one expects \mbox{$c\rightarrow 0$} and the supergravity correction 
vanishes\footnote{The supergravity corrections during radiation domination are
due to K\"{a}hler couplings of the scalar field with the thermal bath 
dominating the density of the Universe. For example, consider a scalar field
$\Psi$, which is part of the thermal bath. Then the supergravity corrections 
arise through the kinetic density due to terms in the K\"{a}hler potential of 
the form: \mbox{$K\sim\Psi^2\Phi^2/m_P^2$}. The kinetic term 
\mbox{${\cal L}_{\rm kin}\equiv
(\partial_m\partial_n K)\partial_\mu\phi_m\partial^\mu\phi_n$}
(with \mbox{$\partial_n\equiv\partial/\partial\phi_n$}) includes a 
contribution of the form \mbox{$\delta{\cal L}_{\rm kin}\sim 
(\Phi/m_P)^2\partial_\mu\Psi\partial^\mu\Psi$}. Now, naively one expects
\mbox{$(\partial\Psi)^2\sim\rho_\Psi\sim T^4$}, because $\Psi$ is part of the 
thermal bath. Since, \mbox{$T^4\sim\rho\sim (Hm_P)^2$} we find that 
\mbox{$\delta{\cal L}_{\rm kin}\sim H^2\Phi^2$}, i.e. supergravity corrections
seem to result again in an effective mass of order $H$. However, a more 
careful examination of the above shows that this is not so. Indeed, 
\mbox{$\partial_\mu\Psi\partial^\mu\Psi=\dot{\Psi}^2-
(\vec\nabla\Psi)^2=0$}, 
because $\Psi$ is a relativistic 
(effectively massless) field, whose modes correspond to plane waves of the 
form: \mbox{$\Psi_k=\Psi_k^0e^{\pm ikt}$}. Similar results are obtained with
fermions. Hence, the supergravity correction vanishes in the radiation 
dominated period. KD wishes to thank T.~Moroi for clarifying this point.}
\cite{sugraRD}.
Now, the reheating temperature is 
\mbox{$T_{\rm reh}\sim\sqrt{\Gamma_{\rm inf}m_P}$}, where $\Gamma_{\rm inf}$ 
is the decay rate of the inflaton field. Using this and Eq.~(\ref{mPhi}) it is 
easy to show that
\begin{equation}
H_{\rm reh}\sim\Gamma_{\rm inf}\geq m_\Phi\quad\Leftrightarrow\quad 
T_{\rm reh}\geq M_S\sim 1\;{\rm TeV}\,,
\label{Treh}
\end{equation}
where the subscript `reh' denotes the time of reheating.
Typically, baryogenesis mechanisms require \mbox{$T_{\rm reh}>1$ TeV}. 
Therefore, reheating occurs before the onset of the quadratic oscillations,
which we discussed in the previous subsection. As a result, after reheating,
the motion of the field is overdamped by the excessive friction of a large
Hubble parameter (compared to its mass) and so $\Phi$ freezes until $H$ 
is reduced enough for the quadratic oscillations to commence. 

To understand this consider the Klein-Gordon equation of motion of the field,
which, in view of Eq.~(\ref{Vsugra}), takes the form
\begin{equation}
\ddot{\Phi}+3H\dot{\Phi}+(\pm cH^2+m_\Phi^2)\Phi=0\,.
\label{KG}
\end{equation}
If, after reheating, $\Phi$ is dominated by its kinetic density 
\mbox{$\rho_{\rm kin}\equiv\frac{1}{2}\dot{\Phi}^2$} then only the first two
terms in the left-hand-side (LHS) of Eq.~(\ref{KG}) are important, which
results in \mbox{$\rho_{\rm kin}\propto a^{-6}$}. Thus, the kinetic density is 
soon depleted away and the field becomes potential density 
dominated.\footnote{We make the conservative assumption that the value of 
$\Phi$ is not dramatically reduced until its density becomes potential 
dominated. The solution of Eq.~(\ref{KG}) is: 
\mbox{$\Phi\simeq\Phi_{\rm reh}-\sqrt{\rho_{\rm kin}^{\rm reh}/\rho_{\rm reh}}
\left(1-\sqrt{H/H_{\rm reh}}\right)\sqrt{6}\,m_P$}. Hence, 
\mbox{$\Phi\simeq\Phi_{\rm reh}$} if 
\mbox{$\rho_{\rm kin}^{\rm reh}\ll\rho_{\rm reh}$}.}
When 
this happens the first term in the LHS of Eq.~(\ref{KG}) becomes negligible.
Then, considering that \mbox{$c=0$} after reheating, it is easy to find
the solution
\begin{equation}
\Phi\simeq\Phi_{\rm reh}\exp\left[-\frac{1}{12}\left(\frac{m_\Phi}{H}\right)^2
\left(1-\frac{H^2}{H_{\rm reh}^2}\right)\right].
\label{freeze}
\end{equation}
From the above it is evident that, in the interval 
\mbox{$m_\Phi<H<H_{\rm reh}$}, the field remains frozen. 
Consequently,
\begin{equation}
\Phi_{\rm osc}\sim\Phi_{\rm reh}\;.
\label{Phioscinf}
\end{equation}
Hence, the required value of $\Phi_{\rm osc}$, given in Eq.~(\ref{Phiosc}),
may be explained by the evolution of $\Phi$ during the period after the end
of inflation until reheating. Below we discuss this evolution assuming
that the sign of the supergravity correction is positive.

The evolution of a scalar field under the influence of supergravity 
corrections has been thoroughly studied in Ref.~\cite{CD}, where it was found
that, during a matter dominated period (such as the one after the end of 
inflation and before reheating, when the Universe is dominated by massive 
inflaton particles), the value of the field is 
given by the following equations:

\bigskip

\noindent
For \mbox{$c>9/16$}:
\begin{equation}
\Phi=\Phi_{\rm inf}
\sqrt{\frac{H}{H_{\rm inf}}}
\left[\cos\!
\left(\sqrt{\frac{16c}{9}-1}\ln\sqrt{\frac{H}{H_{\rm inf}}}
\right)
\!-\!\frac{1}{\sqrt{16c/9-1}}\sin\!
\left(\sqrt{\frac{16c}{9}-1}\ln\sqrt{\frac{H}{H_{\rm inf}}}
\right)
\right]
\end{equation}
For \mbox{$c=9/16$}:
\begin{equation}
\Phi=\Phi_{\rm inf}
\sqrt{\frac{H}{H_{\rm inf}}}
\left(1+\ln\sqrt{\frac{H_{\rm inf}}{H}}\;\right)
\end{equation}
For \mbox{$c<9/16$}:
\begin{eqnarray}
\Phi & = & \Phi_{\rm inf}
\sqrt{\frac{H}{H_{\rm inf}}}
\left[\left(1+\frac{1}{\sqrt{1-16c/9}}\right)
\left(\frac{H_{\rm inf}}{H}\right)^{\frac{1}{2}\sqrt{1-16c/9}}
\right.\nonumber\\
& & \hspace{1.8cm}\left.+\left(1-\frac{1}{\sqrt{1-16c/9}}\right)
\left(\frac{H_{\rm inf}}{H}\right)^{-\frac{1}{2}\sqrt{1-16c/9}}
\;\right],
\end{eqnarray}
which are solutions of Eq.~(\ref{KG}) for \mbox{$m_\Phi^2\ll cH^2$}.
From the above we see that, if \mbox{$\sqrt{c}\leq\frac{3}{4}$}, then the 
field gently rolls toward the origin. On the other hand, if 
\mbox{$\sqrt{c}>\frac{3}{4}$}, then the field oscillates with decreasing 
amplitude and frequency \mbox{$\propto\log\sqrt{H}$}. In all cases, the value 
(or the amplitude) of the field scales as
\begin{equation}
\Phi=\Phi_{\rm inf}\left(\frac{H}{H_{\rm inf}}\right)^{\frac{1}{2}
\left(1-\sqrt{1-16\hat{c}/9}\right)},
\label{Phisugra}
\end{equation}
where 
\begin{equation}
\hat{c}\equiv{\rm min}\{c, 9/16\}\,.
\label{hatc}
\end{equation}
Using Eqs.~(\ref{Phiosc}) and (\ref{Phisugra}) 
[cf. also, Eq.~(\ref{Phioscinf})] one finds
\begin{equation}
\sqrt{\frac{T_{\rm eq}}{M_S}}\sim
\frac{\Phi_{\rm osc}}{m_P}\sim\left(\frac{\Gamma_{\rm inf}}{H_{\rm inf}}
\right)^{\frac{1}{2}\left(1-\sqrt{1-16\hat{c}/9}\right)},
\label{Teq}
\end{equation}
where \mbox{$H_{\rm inf}$} is the Hubble parameter at the end of inflation and
we considered \mbox{$\Phi_{\rm inf}\sim m_P$}. 
From the above, it is easy to obtain
\begin{equation}
\frac{T_{\rm reh}}{V_{\rm inf}^{1/4}}\sim 
10^{-6\left(1-\sqrt{1-16\hat{c}/9}\right)^{-1}},
\label{TrehVinf}
\end{equation}
where we used that \mbox{$\sqrt{T_{\rm eq}/M_S}\sim 10^{-6}$} and also
that \mbox{$V_{\rm inf}^{1/4}\sim\sqrt{H_{\rm inf}m_P}$}
and \mbox{$T_{\rm reh}\sim\sqrt{\Gamma_{\rm inf}m_P}$}.

The amplitude of the density perturbations
(given by the COBE satellite observations), if they are due to the 
amplification of the quantum fluctuations of the inflaton field, 
determines the energy scale of inflation as follows \cite{book}
\begin{equation}
V_{\rm inf}^{1/4}=0.027\epsilon^{1/4}m_P\,,
\label{cobe}
\end{equation}
where $\epsilon$ is one of the, so-called, slow-roll parameters,
associated with the rate of change of $H$ during inflation. Typically,
\mbox{$\epsilon\sim 1/N$} where \mbox{$N\simeq 60$} is the number of the 
remaining e-foldings of inflation when the cosmological scales exit the causal 
horizon. Hence, we see that the energy scale of inflation is determined by
the COBE observations to be given by the energy of grand unification:
\mbox{$V_{\rm inf}^{1/4}\sim 10^{16}$GeV}. 

Inserting this value into Eq.~(\ref{TrehVinf}) we find that, for 
\mbox{$\sqrt{c}>\frac{3}{4}$}, we obtain \mbox{$T_{\rm reh}\sim 10^{10}$GeV}.
A reheating temperature this high is in danger of violating the well-known
gravitino constraint, which requires \mbox{$T_{\rm reh}\leq 10^9$GeV}. 
Enforcing this constraint we find the following allowed range for $c$
\begin{equation}
1\;{\rm TeV}\leq T_{\rm reh}\leq 10^9{\rm GeV}
\quad\Leftrightarrow\quad
\frac{2}{5}\leq c\leq\frac{5}{9}\,.
\label{crange}
\end{equation}
This is a rather narrow range for the value of $c$, albeit quite realistic.
However, this does not necessarily imply any tuning. Indeed, different values 
of $c$ result in different values of $\Phi_{\rm osc}$, which, with $\Phi$ 
being the dark matter, would give different values of $T_{\rm eq}$. The latter 
is determined observationally and has no fundamental origin. Hence, one
can view the above result as an observational determination of $c$.

Still, we can expand the allowed range of $c$ even above 9/16 if we break 
loose from the COBE condition in Eq.~(\ref{cobe}). This is possible if we
consider alternative scenarios for structure formation. For example, if we
assume that the primordial spectrum of density perturbations is due to the
amplification of the quantum fluctuations of some curvaton field {\em other}
than the inflaton, as suggested in Ref.~\cite{curv}, then the COBE constraint
on $V_{\rm inf}^{1/4}$ becomes relaxed into an upper bound \cite{liber}.
Assuming \mbox{$c\geq 9/16$}, Eqs.~(\ref{hatc}) and (\ref{TrehVinf}) give 
%
\begin{equation}
T_{\rm reh}\sim 10^{-6}V_{\rm inf}^{1/4}.
\label{TV}
\end{equation}
Hence, for the allowed range of $T_{\rm reh}$ we find
\begin{equation}
1\;{\rm TeV}\leq T_{\rm reh}\leq 10^9{\rm GeV}
\quad\Leftrightarrow\quad 
10^9{\rm GeV}\leq V_{\rm inf}^{1/4}\leq 10^{15}{\rm GeV}\,.
\label{Vinfrange}
\end{equation}

From the above it is clear that the necessary initial conditions for the 
quadratic oscillations of $\Phi$, in order for the latter to be the dark matter
particle, can be naturally attained by considering the action of supergravity 
corrections on $V(\Phi)$ after the end of inflation and until reheating.

\subsection{\boldmath Avoiding the decay of $\Phi$ into $\phi$-particles}

One final requirement for our dark matter field $\Phi$ is that it should not 
decay into $\phi$-particles until the present time. Indeed, the coupling 
between the two fields suggests that, in principle, such a decay is possible. 
Here we find the appropriate constraint on the coupling constant $\lambda$, 
which ensures that such a decay does not take place.

Let us consider first the perturbative decay of the $\Phi$ condensate. The 
decay rate for the decay: \mbox{$\Phi\rightarrow\phi\;\phi$} is estimated as
%
\begin{equation}
\Gamma_{\Phi\rightarrow\phi\phi}\simeq
\frac{\lambda^2\bar{\Phi}^2}{8\pi m_\Phi}\;.
\label{GPhiphi}
\end{equation}
In order to avoid the decay we need to have 
\mbox{$\Gamma_{\Phi\rightarrow\phi\phi}<H$} until today. Now, since
\mbox{$\bar{\Phi}\propto a^{-3/2}$}, it is easy to find that
\begin{equation}
\frac{\Gamma_{\Phi\rightarrow\phi\phi}}{H}\propto H^{\frac{1-w}{1+w}},
\end{equation}
where $w$ is the barotropic parameter corresponding to the equation of state of
the dominant component of the content of the Universe
(\mbox{$w=0$} \{\mbox{$w=\frac{1}{3}$}\}
for the matter \{radiation\} dominated epoch). Since \mbox{$w\leq 1$}, we see
that the constraint on $\Gamma_{\Phi\rightarrow\phi\phi}$ relaxes with time.
Hence, the tightest constraint corresponds to the earliest time when the decay
\mbox{$\Phi\rightarrow\phi\;\phi$} can occur. Now, this decay is possible only 
when \mbox{$m_\Phi\geq 2m_\phi^{\rm eff}\sim\sqrt{\lambda}\bar{\Phi}$}, where
we considered that, during most of the oscillation period, 
\mbox{$\Phi\sim\bar{\Phi}$}. Hence, the decay can take place only after the
amplitude of the oscillations becomes \mbox{$\bar{\Phi}<\bar{\Phi}_m$}, where
\begin{equation}
\bar{\Phi}_m\equiv\frac{m_\Phi}{\sqrt{\lambda}}\;.
\label{Phim}
\end{equation}
From Eqs.~(\ref{GPhiphi}) and (\ref{Phim}) we find
\begin{equation}
\Gamma_m\equiv\Gamma_{\Phi\rightarrow\phi\phi}(\bar{\Phi}_m)
\sim\lambda m_\Phi\;.
\label{Gm}
\end{equation}
Therefore, the constraint for the avoidance of the decay
\mbox{$\Phi\rightarrow\phi\;\phi$} reeds
\begin{equation}
\Gamma_m<H_m\;,
\label{consdec}
\end{equation}
where \mbox{$H_m\equiv H(\bar{\Phi}=\bar{\Phi}_m)$}.
It can be checked that, in all cases, \mbox{$H_m>H_{\rm eq}$}, which means 
that the amplitude of the $\Phi$--oscillations becomes smaller than 
$\bar{\Phi}_m$ during the radiation epoch, when 
\mbox{$\bar{\Phi}\propto\sqrt{\rho_\Phi}\propto a^{-3/2}\propto H^{3/4}$}. 
Thus, it is easy to find that
\begin{equation}
H_m\sim
H_{\rm eq}\left(\frac{\bar{\Phi}_m}{\bar{\Phi}_{\rm eq}}\right)^{4/3}\sim\;
\lambda^{-2/3}\,\frac{T_{\rm eq}^2}{m_P}
\left(\frac{m_\Phi}{T_{\rm eq}}\right)^{8/3},
\label{Hm}
\end{equation}
where we used that 
\mbox{$m_\Phi\bar{\Phi}_{\rm eq}\sim\sqrt{\rho_{\rm eq}}\sim T_{\rm eq}^2\sim
m_PH_{\rm eq}$}.

Using Eqs.~(\ref{Gm}) and (\ref{Hm}) and enforcing the constraint in 
Eq.~(\ref{consdec}) we obtain
\begin{equation}
\lambda<
\frac{m_\Phi}{m_P}
\left(\frac{m_P}{T_{\rm eq}}\right)^{2/5}
\sim 10^{-19}.
\label{lbound}
\end{equation}

Apart from the perturbative decay of $\Phi$ it is possible that 
$\phi$--particle production occurs in an explosive manner due to parametric 
resonance effects \cite{preh}. This process takes place during the small 
fraction of each oscillation when $\Phi$ is close enough to the origin that 
\mbox{$m_\Phi\geq 2m_\phi^{\rm eff}\simeq 2\sqrt{\lambda}\Phi(t)$} even though
\mbox{$\bar{\Phi}>\bar{\Phi}_m$}. The efficiency of the resonance is 
determined by the so-called $q$--factor:
\begin{equation}
q\sim\frac{\lambda\bar{\Phi}^2}{m_\Phi^2}\sim
\left(\frac{\bar{\Phi}}{\bar{\Phi}_m}\right)^2.
\label{q}
\end{equation}
When \mbox{$q\gg 1$} we are in the broad resonance regime and the
production of $\phi$--particles is quite efficient. However, despite this 
fact, their energy is only a fraction of the total energy in the oscillating 
$\Phi$. Consequently, {\em the evolution of $\Phi$ is hardly affected by the
resonant production of $\phi$--particles.}
The produced $\phi$--particles are expected to eventually thermalize 
and become a (negligible) component of the thermal bath. The resonance becomes 
narrow when \mbox{$q\lsim 1$}, which occurs deep into the radiation epoch. 
Soon afterwards, backreaction and rescattering effects are expected to shut 
down the resonance and terminate the non-perturbative production of 
$\phi$--particles. Hence, the resonant decay of $\Phi$ does not really impose
any additional constraints.\footnote{Note that because of the absence of a 
quartic term $\sim\Phi^4$ in the scalar potential in Eq.~(\ref{V}), we do not 
expect the resonant decay of the zero mode of the oscillating 
$\Phi$--condensate into $\Phi$ bosons of larger momenta \cite{preh}.} 

In view of Eqs.~(\ref{lrange}) and (\ref{lbound}) we see that the allowed range
for $\lambda$ is
\begin{equation}
10^{-30}<\lambda<10^{-19}.
\label{Lrange}
\end{equation}
Such a small coupling between flat directions can be naturally 
realized through being determined by the
Planck suppressed expectation value of some other field.

\section{Dark energy requirements}

\subsection{\boldmath Locking conditions for $\phi$}

Our results also depend on the initial conditions for our quintessence field
$\phi$, which has to find itself near the origin in order to become locked, 
when the $\Phi$ oscillations begin. The required condition, in fact, is
\begin{equation}
\phi_{\rm osc}\ll M\,.
\label{phiosc}
\end{equation}
Recall, here, that the subscript `osc' denotes the onset of the 
oscillations of the field $\Phi$ and not of $\phi$. 

Firstly, due to the interaction term in Eq.~(\ref{V}), it is evident that,
if $V_{\rm inf}^{1/4}$ is small, one may not be able to have both $\Phi$ and 
$\phi$ of the order of $m_P$ after the end of inflation (despite the fact that
\mbox{$\lambda<10^{-19}$} [cf. Eq.~(\ref{Lrange})]) because we need 
\mbox{$V(\Phi,\phi)\ll V_{\rm inf}$}, otherwise the inflationary dynamics 
would be disturbed. As we have chosen \mbox{$\Phi_{\rm inf}\sim m_P$}, we 
obtain the following bound for the value of $\phi$ at the end of 
inflation\footnote{According to this bound and Eq.~(\ref{lrange}) the quartic 
term is \mbox{$\alpha\,\phi_{\rm inf}^4<
(\alpha/\lambda^2)V_{\rm inf}^2/m_P^4\ll V_{\rm inf}$}.}
\begin{equation}
\phi_{\rm inf}\leq\min\left\{\frac{1}{\sqrt{\lambda}}
\left(\frac{V_{\rm inf}^{1/4}}{m_P}\right)^2,1\right\}m_P\;.
\label{phiinf}
\end{equation}
Assuming that $\phi$ is also subject, like $\Phi$, to supergravity 
corrections, which provide a contribution $c'H^2$ to its mass-squared, we can
estimate $\phi_{\rm reh}$ using the analog of Eq.~(\ref{Phisugra})
\begin{equation}
\phi_{\rm reh}=\phi_{\rm inf}
\left(\frac{\Gamma_{\rm inf}}{H_{\rm inf}}\right)^{\frac{1}{2}
\left(1-\sqrt{1-16\hat{c}'/9}\right)},
\label{phisugra}
\end{equation}
where 
\begin{equation}
\hat{c}'\equiv{\rm min}\{c', 9/16\}\,.
\label{hatc'}
\end{equation}
Using Eq.~(\ref{Teq}) it is easy to obtain
\begin{equation}
\phi_{\rm reh}\sim\phi_{\rm inf}
\left(\frac{T_{\rm eq}}{M_S}\right)^{\frac{1}{2}
\left(\frac{1-\sqrt{1-16\hat{c}'/9}}{1-\sqrt{1-16\hat{c}/9}}\right)}.
\label{phireh}
\end{equation}
If both \mbox{$c,c'\geq 9/16$} then the above gives 
\mbox{$\phi_{\rm reh}\sim 10^{-6}\phi_{\rm inf}$}. However, one can achieve
a substantially smaller $\phi_{\rm reh}$. For example, with 
\mbox{$c'\geq 9/16$} and \mbox{$c\approx 0.4$} one finds 
\mbox{$\phi_{\rm reh}^{\rm min}\sim 10^{-13}\phi_{\rm inf}$}.

As in the case of $\Phi$, the supergravity corrections disappear (they cancel
out) after reheating. Consequently, during the radiation dominated epoch, 
the effective mass of $\phi$, according to
Eq.~(\ref{mphi}), is given by
\begin{equation}
m_\phi^{\rm eff}\sim\sqrt{\lambda}\,\Phi_{\rm osc}\,,
\label{meff}
\end{equation}
where we considered that \mbox{$\Phi_{\rm reh}\sim\Phi_{\rm osc}\gg\Phi_c$}
[cf. Eq.~(\ref{Phioscinf})]. Since, in the interval
\mbox{$m_\Phi<H<\Gamma_{\rm inf}$}, $\Phi$ remains frozen,
the above effective mass remains constant
after reheating and until the oscillations of $\Phi$ begin. Comparing this
effective mass with $\Gamma_{\rm inf}$ one finds
\begin{equation}
m_\phi^{\rm eff}>\Gamma_{\rm inf}\quad\Leftrightarrow\quad
T_{\rm reh}<\lambda^{1/4}10^{-3}m_P\;,
\end{equation}
where we also used Eq.~(\ref{Teq}). In view of Eq.~(\ref{lrange}) we find
\mbox{$\lambda^{1/4}10^{-3}m_P>10^8$GeV}. Hence, considering the 
gravitino constraint \mbox{$T_{\rm reh}\leq 10^9$GeV}, we expect that 
\mbox{$m_\phi^{\rm eff}>\Gamma_{\rm inf}$} and, therefore, the oscillations
of $\phi$ begin immediately after reheating.

During these oscillations we have 
\mbox{$\phi\propto\sqrt{\rho_\phi}\propto a^{-3/2}\propto H^{3/4}$}, which
results in
\begin{equation}
\phi_{\rm osc}\sim
\phi_{\rm reh}\left(\frac{m_\Phi}{\Gamma_{\rm inf}}\right)^{3/4}
\sim\phi_{\rm reh}\left(\frac{M_S}{T_{\rm reh}}\right)^{3/2}.
\label{phioscreh}
\end{equation}
For the allowed range of $T_{\rm reh}$ the above corresponds to
\mbox{$10^{-9}\leq\phi_{\rm osc}/\phi_{\rm reh}\leq 1$}.
%

Putting Eqs.~(\ref{phiinf}), (\ref{phireh}) and (\ref{phioscreh}) together
we obtain:
\begin{equation}
\phi_{\rm osc}\leq\left(\frac{M_S}{T_{\rm reh}}\right)^{3/2}
\left(\frac{T_{\rm eq}}{M_S}\right)^{\frac{1}{2}
\left(\frac{1-\sqrt{1-16\hat{c}'/9}}{1-\sqrt{1-16\hat{c}/9}}\right)}
\min\left\{\frac{1}{\sqrt{\lambda}}
\left(\frac{V_{\rm inf}^{1/4}}{m_P}\right)^2,1\right\}m_P\;.
\label{phiresult}
\end{equation}
The first factor on the right-hand-side of the
above can be as low as $10^{-9}$, the second one can
be as low as $10^{-13}$, while the last factor in front of $m_P$
cannot be larger than unity. Hence, it is evident that the requirement
in Eq.~(\ref{phiosc}), which demands \mbox{$\phi_{\rm osc}<10^{-15}m_P$},
may well be satisfied.

Let us demonstrate this with a small example. Suppose that that 
\mbox{$V_{\rm inf}^{1/4}\sim 10^{16}$GeV} and we choose
\mbox{$c\approx 0.4$} and \mbox{$c'\geq 9/16$}. 
Using this and in view also of 
Eq.~(\ref{Lrange}), Eq.~(\ref{phiinf}) suggests that 
\mbox{$\phi_{\rm inf}\lsim m_P$}. Hence, Eq.~(\ref{phiresult}) suggests that
\mbox{$\phi_{\rm osc}<10^{-15}m_P$} can be achieved if 
\mbox{$T_{\rm reh}\geq 10\,M_S$}, which allows almost the entire range of 
$T_{\rm reh}$.

As a result of the above, our assumption \mbox{$\phi\simeq 0$} in 
Sec.~\ref{model} is well justified.

\subsection{\boldmath The mass and VEV of $\phi$}\label{VEV}

In order to achieve a false vacuum density as small as $\rho_0$ we not only
require a small tachyonic mass for our locked quintessence field $\phi$ but
also a small VEV according to Eq.~(\ref{choice}). One way to achieve this is 
to stabilize the $\phi$--direction by means of some high-order 
non-renormalizable term, of the form
\begin{equation}
V(\Phi=0,\phi)=V_0-\frac{1}{2}m_\phi^2\phi^2+\frac{\phi^{2n}}{Q^{2n-4}},
\label{Vnonren}
\end{equation}
where \mbox{$n>2$} and $Q$ is an appropriate large cut-off scale, which is
linked to the VEV $M$ as
\begin{equation}
Q\sim\alpha^{-\frac{1}{2(n-2)}}M\,,
\label{QM}
\end{equation}
where we also considered Eq.~(\ref{mphi0}).
The most natural choice is \mbox{$Q=m_P$}, which gives \mbox{$n=4$}.
Hence, the action of non-renormalizable terms may well reduce the VEV of 
$\phi$ naturally\footnote{The absence of a quartic term can be understood in a 
similar manner as with the so-called flaton fields discussed in 
Ref~\cite{flaton}.}.

The important issue here is that we need to preserve the smallness of the 
tachyonic mass. For a flat direction one expects the dominant contribution 
to the mass to be of the form \mbox{$(M_S/m_P)^pM_S$} with \mbox{$p=1$}. 
This is the case, for example, of the dark matter field $\Phi$ 
as shown in Eq.~(\ref{mPhi}). However, in the case of $\phi$, we need to 
suppress this contribution and consider \mbox{$p=2$} instead,
according to Eq.~(\ref{mphiMS}). It is conceivable that this may occur due
to accidental cancellations in the K\"{a}hler potential, or due to some 
symmetry, which protects $m_\phi$. In any case, even if this requirement 
corresponds to a certain level of fine-tuning, this tuning is much less 
stringent than what is required in most quintessence models (with typical 
effective mass \mbox{$m_Q\sim H_0$}), because \mbox{$m_\phi\sim 10^{15} H_0$}. 
Moreover, since \mbox{$m_\phi\sim 10^9H_{\rm eq}$}, 
supergravity corrections, during the matter era after $t_{\rm eq}$, are 
negligible, in contrast to the usual quintessence models \cite{KL}.
Note, however, that there exist some dark energy models corresponding to 
particles with mass much larger than $H_0$. For example this is possible in 
scalar tensor theories of gravity \cite{ST}, which can account for both 
quintessence and dark matter (e.g. see Refs.~\cite{STQ} and \cite{STDM} 
respectively). Another recent such example is dark energy from mass varying 
neutrinos \cite{neutrin}.

A marginal increase of $m_\phi$ may be achieved if we consider that
the VEV of $\phi$ is reduced by the action of loop corrections (instead 
of non-renormalizable terms). These are of the form
\begin{equation}
V(\Phi=0,\phi)=V_0-\frac{1}{2}m_\phi^2\phi^2+Cm_\phi^2\ln(\phi/Q)\phi^2,
\label{Vloop}
\end{equation}
where \mbox{$C\ll 1$}. In this case we have
\begin{equation}
M\sim\exp(-1/2C)Q\,.
\label{MQ}
\end{equation}
The above setup can increase the tachyonic mass by a factor $1/\sqrt{C}$.
The best case, however, corresponds to \mbox{$Q=m_P$}, which gives
\mbox{$C\simeq 0.015$}, i.e. $m_\phi$ is increased at most by a factor of~8.

Finally, the smallness of the tachyonic mass of our locked quintessence may 
result in the appearance of a fifth-force \cite{5th} because the associated 
Compton wavelength is
\begin{equation}
\ell_\phi\sim m_\phi^{-1}\sim(10^{-27}{\rm GeV})^{-1}\sim 1\;{\rm A.U.}
\label{5th}
\end{equation}
From the above we see that such a fifth-force cannot bias the formation of
large structures like galaxies and galactic clusters. It is conceivable, 
though, that it may affect the generation of population~III stars and stellar 
formation in general. However, the fifth-force is strongly constrained by the 
solar system tests on the equivalence principle \cite{equiv}. 
Hence, we require that $\phi$ is some hidden sector 
field with suppressed interactions with ordinary baryonic matter.

\section{Discussion and conclusions}

We have analyzed a unified model for the dark matter and the dark energy. As 
dark matter we used a modulus field $\Phi$, which corresponds to a flat 
direction of supersymmetry. The field undergoes coherent oscillations that
correspond to massive particles (WIMPs), constituting pressureless matter. Our 
$\Phi$ field is weakly coupled with another scalar $\phi$, through a hybrid 
type potential, very common in supersymmetric theories. The scalar $\phi$ 
corresponds to a flat direction lifted by 
non-renormalizable terms. Due to the above coupling the oscillating $\Phi$ 
keeps $\phi$ `locked' on top of the saddle point of the potential, resulting 
in a non-zero false vacuum contribution $V_0$. The amplitude of the 
oscillations decreases in time due to the Universe expansion. Below a certain 
value $\Phi_\Lambda$ the Universe becomes dominated by the false vacuum 
density and a phase of accelerated expansion begins. Acceleration continues 
until the amplitude of the oscillations decreases down to a critical value 
$\Phi_c$, when the `locked' quintessence field $\phi$ is released and rolls 
down to its VEV. At this point the system reaches the true vacuum and the 
accelerated expansion ceases. 

We have shown that it is possible to explain both dark matter and dark energy
by taking the supersymmetry breaking scale to be \mbox{$M_S\sim$ TeV},
which corresponds to low-scale gauge-mediated supersymmetry breaking. The VEV 
of $\phi$ has to be given also by $M_S$, which is possible to achieve by 
stabilizing its potential with the use of a non-renormalizable term of the 
8th order. Hence, using only two natural mass scales, $m_P$ and 
\mbox{$M_S\sim m_{3/2}$} we are able to achieve cosmic coincidence, in the 
sense that we manage to obtain comparable densities for dark matter and dark 
energy at present without severe fine--tuning.
In order to successfully account for both dark matter and dark energy our 
scalar fields need to have the correct initial conditions. By studying the
dynamics of our scalar fields in the early Universe, we have demonstrated
that the required initial conditions are naturally attained when considering 
the action of supergravity corrections to the scalar potential during the 
period following the end of primordial inflation and until reheating.

The advantages of our model are the following. Firstly, it uses a 
theoretically well motivated framework to address, in a unified manner, both 
the open issues of dark matter and dark energy. Also, coincidence is achieved 
with the use of only natural energy scales and initial conditions. 
The observational consequences of our 
model are similar to those of $\Lambda$CDM because, during most of the 
evolution of the Universe, the model is reduced to Eq.~(\ref{over}) (i.e. it 
corresponds to a collection of $\Phi$--particles (WIMPs) plus an effective 
cosmological constant \mbox{$\Lambda_{\rm eff}=V_0/m_P^2$}). Hence, our model 
enjoys all the successes of $\Lambda$CDM but it does not suffer from its 
disadvantages, namely the extreme fine-tuning of $\Lambda$ and also the 
conceptual blunders of eternal acceleration and future causal horizons. Our 
model avoids eternal acceleration because the `locked' quintessence field 
terminates false vacuum domination, when it is released from the origin. The 
only tuning problem that our model suffers from is the smallness of the 
tachyonic mass $m_\phi$ of our quintessence field, which may be due to some 
approximate symmetry. Still, we have \mbox{$m_\phi\gg H_{\rm eq}\gg H_0$}, 
which means that the supergravity corrections to the $\phi$-direction are 
negligible even after $t_{\rm eq}$, in contrast to the generic problem of most 
quintessence models, which have \mbox{$m_Q\sim H_0$}. 

It is interesting to estimate how long the late period of accelerated 
expansion lasts. After domination by the false vacuum we have 
\mbox{$H_0\simeq\sqrt{V_0}/\sqrt{3}\,m_P=$ constant}. Hence, a phase of 
(quasi) de~Sitter expansion begins, with \mbox{$a\simeq a_0\exp(H_0\Delta t)$},
where \mbox{$\Delta t=t-t_0$}. Now, for the oscillating $\Phi$ we have
\mbox{$\Phi\propto\sqrt{\rho_\Phi}\propto a^{-3/2}$}. Thus, we obtain
\begin{eqnarray}
 & \Phi_c\simeq\Phi_\Lambda\exp(-\frac{3}{2}H_0\Delta t_c) & 
\Rightarrow\qquad
\Delta t_c\simeq\frac{2}{3}
\left[\,\ln\left(\frac{m_P}{M_S}\right)
+\ln\sqrt{\lambda}\,\right]H_0^{-1},
\end{eqnarray}
where we have used Eqs.~(\ref{Phic}), (\ref{PhiL}) and (\ref{mPhi}). 
In view of Eqs.~(\ref{choice}) and (\ref{Lrange}) we see that the period of
acceleration may last up to 8 Hubble times (e-foldings) depending on the 
value of $\lambda$. 
Another interesting point regards the coupling $g_\Phi$ of the dark matter 
particle to its decay products. Eqs.~(\ref{GPhi}), (\ref{mPhi}) and 
(\ref{choice}) suggest that $g_\Phi$ should lie in the range: 
\mbox{$10^{-30}\leq g_\Phi<10^{-15}$}, with 
the lower bound corresponding to gravitational decay, when 
\mbox{$g_\Phi\sim m_\Phi/m_P$}.
Thus, $\Phi$ is truly a WIMP [cf. also Eq.~(\ref{Lrange})].

We should also point out here that our oscillating $\Phi$--condensate does not 
have to be the dark matter necessarily. Indeed, it is quite possible that 
$\phi$--remains locked on top of the false vacuum while $\rho_\Phi$ is 
negligible at present. That way, our model can account for the dark energy
requirements even if the initial conditions for $\Phi$ (e.g. the value of $c$) 
are not appropriate for the latter to be the dark matter. Indeed, the locking
of quintessence requires that \mbox{$\rho_\Phi(t_0)\geq\rho_\Phi^{\rm min}$},
where \mbox{$\rho_\Phi^{\rm min}\sim m_\Phi^2\Phi_c^2$} corresponds to the 
minimum energy for the oscillations. In view of Eqs.~(\ref{Phic}), (\ref{r0}),
(\ref{mPhi}) and (\ref{choice}) it is easy to find:
%
\mbox{$\rho_\Phi^{\rm min}/\rho_0\sim 10^{-30}\lambda^{-1}$}.
Hence, depending on $\lambda$, $\Phi$ may contribute only by a small 
fraction to dark matter, while still being able to lock quintessence and
cause the observed accelerated expansion at present. However, we feel that 
using $\Phi$ to account also for the dark matter renders our model much more
effective and economical, without any additional tuning requirements (in the 
sense that the required value of $c$ is natural).

To summarize we have presented a unified model of dark matter and dark energy
in the context of low-scale gauge-mediated supersymmetry breaking. Our model
retains the predictions of $\Lambda$CDM, while avoiding eternal acceleration 
and achieving coincidence without significant fine-tuning. The initial 
conditions of our model are naturally attained due to the effect of 
supergravity corrections to the scalar potential in the early Universe,
following a period of primordial inflation. 

\bigskip

\noindent
{\Large\bf Acknowledgments}

\medskip

\noindent
We are grateful to the journal referee for his/her valuable comments
and constructive criticism.


\begin{thebibliography}{99}

\bibitem{wmap}
D.~N.~Spergel {\it et al.},
Astrophys.\ J.\ Suppl.\  {\bf 148}, 175 (2003).

\bibitem{sdss}
M.~Tegmark {\it et al.}  [SDSS Collaboration],
astro-ph/0310723.

\bibitem{2dF}
M.~Colless,
astro-ph/0305051.

\bibitem{SN}
S.~Perlmutter {\it et al.}  [Supernova Cosmology Project Collaboration],
Astrophys.\ J.\  {\bf 517}, 565 (1999);
A.~G.~Riess {\it et al.}  [Supernova Search Team Collaboration],
Astron.\ J.\  {\bf 116}, 1009 (1998).

\bibitem{review}
P.~J.~E.~Peebles and B.~Ratra,
Rev.\ Mod.\ Phys.\  {\bf 75}, 559 (2003).

\bibitem{horizons}
S.~Hellerman, N.~Kaloper and L.~Susskind,
JHEP {\bf 0106}, 003 (2001);
W.~Fischler, A.~Kashani-Poor, R.~McNees and S.~Paban,
JHEP {\bf 0107}, 003 (2001);
E.~Witten,
hep-th/0106109;
N.~Goheer, M.~Kleban and L.~Susskind,
JHEP {\bf 0307}, 056 (2003).

\bibitem{quint}
L.~M.~Wang, R.~R.~Caldwell, J.~P.~Ostriker and P.~J.~Steinhardt,
Astrophys.\ J.\  {\bf 530}, 17 (2000);
I.~Zlatev, L.~M.~Wang and P.~J.~Steinhardt,
Phys.\ Rev.\ Lett.\  {\bf 82}, 896 (1999);
G.~Huey, L.~M.~Wang, R.~Dave, R.~R.~Caldwell and P.~J.~Steinhardt,
Phys.\ Rev.\ D {\bf 59}, 063005 (1999);
R.~R.~Caldwell, R.~Dave and P.~J.~Steinhardt,
Phys.\ Rev.\ Lett.\  {\bf 80}, 1582 (1998).

\bibitem{KL}
C.~F.~Kolda and D.~H.~Lyth,
Phys.\ Lett.\ B {\bf 458}, 197 (1999).

\bibitem{other}
R.~Bean and J.~Magueijo,
Phys.\ Lett.\ B {\bf 517}, 177 (2001);
L.~Amendola and D.~Tocchini-Valentini,
Phys.\ Rev.\ D {\bf 64} (2001) 043509;
D.~Tocchini-Valentini and L.~Amendola,
Phys.\ Rev.\ D {\bf 65}, 063508 (2002);
W.~Zimdahl and D.~Pavon,
Phys.\ Lett.\ B {\bf 521}, 133 (2001);
M.~Pietroni,
Phys.\ Rev.\ D {\bf 67}, 103523 (2003);
T.~Padmanabhan and T.~R.~Choudhury,
Phys.\ Rev.\ D {\bf 66}, 081301 (2002);
D.~Comelli, M.~Pietroni and A.~Riotto,
Phys.\ Lett.\ B {\bf 571}, 115 (2003);
H.~Ziaeepour,
Phys.\ Rev.\ D {\bf 69} (2004) 063512;
G.~R.~Farrar and P.~J.~E.~Peebles,
astro-ph/0307316;
E.~I.~Guendelman and A.~B.~Kaganovich,
gr-qc/0312006.

\bibitem{dvali}
G.~Dvali and S.~Kachru,
hep-th/0309095;
hep-ph/0310244.

\bibitem{ours}
M.~Axenides and K.~Dimopoulos,
hep-ph/0310194.

\bibitem{turner}
M.~S.~Turner,
Phys.\ Rev.\ D {\bf 28}, 1243 (1983).

\bibitem{sugra}
M.~Dine, L.~Randall and S.~Thomas,
Nucl.\ Phys.\ B {\bf 458}, 291 (1996);
Phys.\ Rev.\ Lett.\  {\bf 75}, 398 (1995).

\bibitem{sugraRD}
D.~H.~Lyth and T.~Moroi,
JHEP {\bf 0405} (2004) 004.

\bibitem{CD}
K.~Dimopoulos, G.~Lazarides, D.~Lyth and R.~Ruiz de Austri,
Phys.\ Rev.\ D {\bf 68}, 123515 (2003).

\bibitem{book}
A.~R.~Liddle and D.~H.~Lyth,
{\it Cosmological inflation and large-scale structure}
(Cambridge Univ. Press, Cambridge U.K., 2000).

\bibitem{curv}
D.~H.~Lyth and D.~Wands,
Phys.\ Lett.\ B {\bf 524}, 5 (2002);
D.~H.~Lyth, C.~Ungarelli and D.~Wands,
Phys.\ Rev.\ D {\bf 67}, 023503 (2003);
T.~Moroi and T.~Takahashi,
Phys.\ Rev.\ D {\bf 66}, 063501 (2002);
K.~Enqvist, S.~Kasuya and A.~Mazumdar,
Phys.\ Rev.\ Lett.\  {\bf 90}, 091302 (2003).

\bibitem{liber}
K.~Dimopoulos and D.~H.~Lyth,
hep-ph/0209180.

\bibitem{preh}
L.~Kofman, A.~D.~Linde and A.~A.~Starobinsky,
Phys.\ Rev.\ D {\bf 56} (1997) 3258.

\bibitem{flaton}
D.~H.~Lyth and E.~D.~Stewart,
Phys.\ Rev.\ D {\bf 53} (1996) 1784.

\bibitem{ST}
P.~P.~Fiziev,
gr-qc/0202074.

\bibitem{STQ}
P.~P.~Fiziev and D.~A.~Georgieva,
Phys.\ Rev.\ D {\bf 67} (2003) 064016;
D.~F.~Torres,
Phys.\ Rev.\ D {\bf 66} (2002) 043522;
L.~M.~Diaz-Rivera and L.~O.~Pimentel,
Int.\ J.\ Mod.\ Phys.\ A {\bf 18} (2003) 651;
A.~Bhadra and K.~K.~Nandi,
gr-qc/0203103;
E.~Elizalde, S.~Nojiri and S.~D.~Odintsov,
hep-th/0405034.

\bibitem{STDM}
J.~A.~Casas, J.~Garcia-Bellido and M.~Quiros,
Class.\ Quant.\ Grav.\  {\bf 9} (1992) 1371,
J.~Garcia-Bellido,
Int.\ J.\ Mod.\ Phys.\ D {\bf 2} (1993) 85.

\bibitem{neutrin}
R.~Fardon, A.~E.~Nelson and N.~Weiner,
astro-ph/0309800;
D.~B.~Kaplan, A.~E.~Nelson and N.~Weiner,
hep-ph/0401099.


\bibitem{5th}
P.~Fayet,
hep-ph/0111282;
T.~Damour,
gr-qc/0109063.

\bibitem{equiv}
C.M.~Will, {\it Theory and Experiment in Gravitiational Physics}
(Basic Books/Perseus Group, New York, 1993);
C.~M.~Will,
Living Rev.\ Rel.\  {\bf 4}, 4 (2001).
C.~Talmadge, J.~P.~Berthias, R.~W.~Hellings and E.~M.~Standish,
Phys.\ Rev.\ Lett.\  {\bf 61}, 1159 (1988);
J.O.~Dickey {\it et al.}, Science {\bf 265}, 482 (1994).

\end{thebibliography}
\end{document}